\documentclass[%
 reprint,
 amsmath,amssymb,
 aps,
]{revtex4-2}

\usepackage{color}
\usepackage{graphicx}
\usepackage{subfigure}
\usepackage[utf8]{inputenc}
\usepackage{dcolumn}
\usepackage{bm}
\usepackage{mathrsfs}
\usepackage{tipa}

\begin{document}

\preprint{APS/123-QED}

\title{Burst-tree structure and higher-order temporal correlations}

\author{Tibebe Birhanu}
\affiliation{Department of Physics, The Catholic University of Korea, Bucheon 14662, Republic of Korea}

\author{Hang-Hyun Jo}
\email{h2jo@catholic.ac.kr}
\affiliation{Department of Physics, The Catholic University of Korea, Bucheon 14662, Republic of Korea}
\date{\today}

\begin{abstract}
Understanding characteristics of temporal correlations in time series is crucial for developing accurate models in natural and social sciences. The burst-tree decomposition method was recently introduced to reveal temporal correlations in time series in a form of an event sequence, in particular, the hierarchical structure of bursty trains of events for the entire range of timescales [Jo et al., Sci.~Rep.~\textbf{10}, 12202 (2020)]. Such structure cannot be solely captured by the interevent time distribution but can show higher-order correlations beyond interevent times. It has been found to be simply characterized by the burst-merging kernel governing which bursts are merged together as the timescale for defining bursts increases. In this work, we study the effects of kernels on the higher-order temporal correlations in terms of burst size distributions, memory coefficients for bursts, and the autocorrelation function. We employ several kernels, including the constant, sum, product, and diagonal kernels as well as those inspired by empirical results. We generically find that kernels with preferential merging lead to heavy-tailed burst size distributions, while kernels with assortative merging lead to positive correlations between burst sizes. The decaying exponent of the autocorrelation function depends not only on the kernel but also on the power-law exponent of the interevent time distribution. In addition, thanks to the analogy to the coagulation process, analytical solutions of burst size distributions for some kernels could be obtained. Our findings may shed light on the role of burst-merging kernels as underlying mechanisms of higher-order temporal correlations in time series.
\end{abstract}

\maketitle


\section{Introduction}

Complex systems, which are ubiquitous in both natural and social systems, are distinguished by their intricate networks of interactions between individual components, leading to emergent complex dynamics~\cite{Barabasi2016Network, Newman2018Networks, Menczer2020First, Dorogovtsev2022Nature, Holme2012Temporal, Holme2023Temporal, Masuda2016Guide}. These systems often exhibit bursty behaviors, where periods of intense activity are interspersed with periods of quiescence~\cite{Barabasi2005Origin, Karsai2018Bursty}. Bursty dynamics have been observed in a wide range of systems, including physical~\cite{Bak1987Selforganized, Jensen1998Selforganized, Wheatland1998WaitingTime, Corral2004Longterm}, biological~\cite{Beggs2003Neuronal, Petermann2009Spontaneous, Kemuriyama2010Powerlaw}, and social systems~\cite{Goh2008Burstiness, Crane2008Robust, Rybski2009Scaling, Rocha2010Information, Aoki2016Inputoutput, Harang2017Burstiness, Karsai2018Bursty, Jo2020Bursttree, Rocha2020Dynamic, Choi2021Individualdriven}. Therefore, understanding temporal correlations in bursty time series of various complex systems is crucial for developing accurate models in natural and social sciences. 

Temporal correlations of time series in the form of event sequences have been studied in terms of an interevent time (IET), a bursty train, and an autocorrelation function (ACF) among others. The IET is defined as a time interval between two consecutive events, and its distribution often shows a heavy tail in various complex systems~\cite{Karsai2018Bursty, Goh2008Burstiness}. The notion of bursty trains or bursts was introduced to detect the correlations between an arbitrary number of consecutive events or consecutive IETs~\cite{Karsai2012Universal}. A burst is a cluster of rapidly occurring events; for a given timescale $\Delta t$, consecutive events in the same burst are separated by IETs less than or equal to $\Delta t$, whereas IETs between events in different bursts are larger than $\Delta t$. The number of events in each burst is called a burst size and its distributions are found to be heavy tailed in several empirical datasets~\cite{Karsai2012Universal, Karsai2012Correlated, Yasseri2012Dynamics, Jiang2013Calling, Kikas2013Bursty, Wang2015Temporal, Jo2020Bursttree}. Finally, the ACF is one of the most commonly used methods detecting temporal correlations~\cite{Fano1950Shorttime, Kantelhardt2001Detecting, Jo2024Temporal}. The ACF tends to show a power-law decaying behavior in the presence of long-term memory effects~\cite{Karsai2012Universal, Panzarasa2015Emergence}. We remark that these measures are closely related to each other~\cite{Jo2023Bursty, Jo2024Temporal}.

Recently, Jo et al.~introduced a novel method called burst-tree decomposition to analyze temporal correlations in the event sequence~\cite{Jo2020Bursttree}. This method represents an event sequence in terms of the burst tree, which is decomposed into a set of IETs and an ordinal burst tree. The ordinal burst tree exactly captures the hierarchical structure of correlations between IETs in a systematic way. It turns out that several time series data from diverse backgrounds with different IET distributions show the similar burst-tree structure in terms of heavy-tailed burst size distributions and positive correlations between consecutive burst sizes for a wide range of timescales. The latter was quantified in terms of a memory coefficient for bursts; note that the memory coefficient has been originally proposed for detecting correlations between two consecutive IETs~\cite{Goh2008Burstiness}. Each node of the burst tree indicates a merge of two consecutive bursts that happens when the timescale $\Delta t$ exceeds the IET separating these two bursts. Which bursts are merged together as the timescale increases could be simply summarized by a function of burst sizes of two merged bursts, which is called a burst-merging kernel. Burst-merging kernels estimated from several empirical time series have revealed the preferential and assortative mixing structure of bursts in the data~\cite{Jo2020Bursttree}. 

The burst-merging kernel can be interpreted as a selection rule of two bursts to be merged when generating the burst tree from the individual events. Once a burst-merging kernel is given either from the data or as a simple mathematical function, one can generate a burst tree, hence an event sequence. Indeed, a simplified version of empirically estimated kernels has been shown to successfully generate higher-order temporal correlations observed in the original time series~\cite{Jo2020Bursttree}. Yet, analytical derivation of higher-order temporal correlations from the given kernel has been overwhelmingly difficult. Therefore, we take an alternative approach by studying the simplest, analytically tractable kernels first and then by considering the more complicated, realistic forms of kernels focused on preferential and assortative merging of bursts. By doing so, one can deepen the understanding of the effects of different features of the kernels on the higher-order temporal correlations in the generated time series. We remark that application of our method to the empirical data is beyond the scope of our current work.

The modeling approach using burst-merging kernels is analogous to the coagulation process by which unit particles are merged according to some selection rules or kernels, to become clusters~\cite{Stockmayer1943Theory, Lushnikov1973Evolution, White1982Form, Hendriks1983Coagulation, Aldous1999Deterministic, Lee2001Survey, Leyvraz2003Scaling, Leyvraz2005Rigorous, Wattis2006Introduction}. Different approaches have been taken to derive cluster size distributions as well as power-law exponent values in the cases of power-law distributions of cluster sizes. In particular, a discrete deterministic mean-field approach was employed to derive the analytical solution of the cluster size distribution for three solvable kernels, namely, constant, sum, and product kernels~\cite{Wattis2006Introduction}. Similar results for the diagonal kernel have been obtained in Ref.~\cite{Leyvraz2005Rigorous}. Since unit particles and clusters in the coagulation process respectively correspond to individual events and bursts in the event sequence, the analytical solutions of cluster size distributions in the coagulation process can be translated into the burst size distributions of the event sequence generated using the same kernels. It implies that the analytical solutions of cluster size distributions in Refs.~\cite{Wattis2006Introduction} and~\cite{Leyvraz2005Rigorous} can be used to confirm the burst size distributions obtained when the constant, sum, product, and diagonal kernels are employed as the burst-merging kernels.

Despite the importance of burst-merging kernels in time series analysis, their impact on higher-order temporal correlations beyond the IET distribution remains poorly understood. To address this issue, we conduct a systematic comparison of various kernels, including the constant, sum, product, and diagonal kernels as well as those inspired by empirical results. By analyzing the event sequences generated using these kernels and power-law IET distributions as inputs, we investigate burst size distributions, memory coefficients for bursts, and ACFs. We generically find that kernels with preferential merging lead to the heavy-tailed burst size distributions, while kernels with assortative merging lead to positive correlations between burst sizes. The decaying exponent of the autocorrelation function depends not only on the kernel but also on the power-law exponent of the IET distribution. These findings provide new insights into the role of burst-merging kernels as underlying mechanisms of higher-order temporal correlations in empirical time series.

The paper is organized as follows. We briefly introduce the burst-tree decomposition method in Sec.~\ref{sec:burst_tree}. In Sec.~\ref{sec:model}, we define the models for generating event sequences using various burst-merging kernels and introduce measures for characterizing higher-order temporal correlations. Then, we present the simulation results in Sec.~\ref{sec:result}. Finally, we conclude the paper by summarizing the key findings and their implications in Sec.~\ref{sec:conclusion}.

\section{Burst-tree decomposition method}\label{sec:burst_tree}

The burst-tree decomposition method was recently introduced to systematically reveal temporal correlations in time series, in particular, the hierarchical structure of bursts for a wide range of timescales defining the bursts~\cite{Jo2020Bursttree}. A burst refers to a sequence of events that occur in rapid succession, followed by a period of relative inactivity~\cite{Barabasi2005Origin}. To be precise, for a given timescale $\Delta t$, a burst is defined as a set of events where interevent times (IETs) between consecutive events in the burst are less than or equal to $\Delta t$, whereas the IETs between bursts are greater than $\Delta t$~\cite{Karsai2012Universal}. It implies that two consecutive bursts detected using a timescale $\Delta t$ would be merged for a bigger timescale $\Delta t'$ ($>\Delta t$) if $\Delta t'$ is larger than the IET separating these two bursts. Such merging pattern is exactly captured by the burst-tree decomposition method.

\begin{figure}[!t]
\centering
\includegraphics[width=\columnwidth]{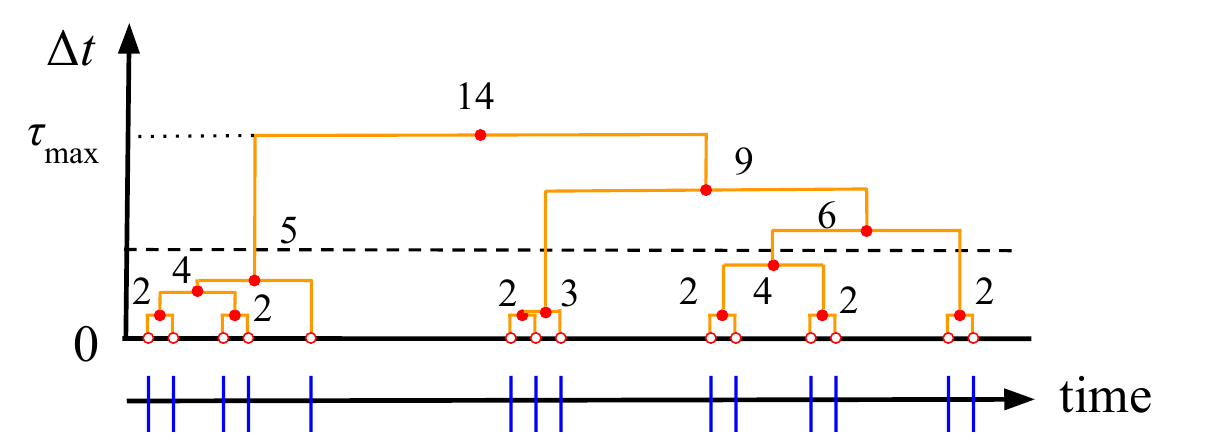}
\caption{Schematic diagram of the burst-tree decomposition method. Vertical lines on the horizontal time axis represent events. The vertical axis in the upper panel represents the timescale $\Delta t$ ranging from $0$ to $\tau_{\rm max}$. For $\Delta t < \tau_{\rm min}$, each event is a leaf node denoted by a red empty circle. As $\Delta t$ increases, consecutive bursts merge to become larger bursts, denoted by red filled circles. The numbers next to the circles are burst sizes. When $\Delta t=\tau_{\rm max}$, all events form a giant burst. The dashed horizontal line at a certain $\Delta t$ yields a sequence of burst sizes, which is $\{5, 3, 4, 2\}$.}
\label{fig:burst_tree_fig}
\end{figure}

Let us consider an event sequence of $n$ events, i.e., $\mathcal{E}=\{t_0,\ldots,t_{n-1}\}$, where $t_i$ is the timing of the $i$th event. From $\mathcal{E}$, one can get the IET sequence as $\{\tau_1,\ldots,\tau_{n-1}\}$, where $\tau_i\equiv t_i-t_{i-1}$. The minimum and maximum IETs are obtained from the IET sequence and then denoted by $\tau_{\rm min}$ and $\tau_{\rm max}$, respectively. To apply the burst-tree decomposition method as depicted in Fig.~\ref{fig:burst_tree_fig}, we start by setting $\Delta t$ to be smaller than $\tau_{\rm min}$. Then each event, as shown by a red empty circle in Fig.~\ref{fig:burst_tree_fig}, is considered to be a burst of size one, implying that each event makes an individual burst. As $\Delta t$ continuously increases to $\tau_{\rm max}$, the bursts start merging with each other to form bigger bursts. Each merge is depicted as a red filled circle in Fig.~\ref{fig:burst_tree_fig}. When $\Delta t= \tau_{\text{max}}$, all events in the event sequence belong to a single giant burst. This merging process with an increasing $\Delta t$ can be visualized as a rooted tree; leaf nodes represent individual events, internal nodes represent merged bursts, and the root node represents a giant burst containing all the events. Therefore, the burst-tree decomposition method provides a hierarchical representation of the event sequence, revealing the underlying structure of the bursty behavior without loss of information in the original event sequence.

Here we assume that each merge occurs only between two bursts. Despite the fact that more than two bursts can merge at the same time if IETs separating them are identical, such merge rarely happens for a broad distribution of IETs~\cite{Jo2020Bursttree}. Then, each internal node, indexed by $u$, is a parent node of two children nodes; the left (right) child node is indexed by $v$ ($w$). The burst size associated with the node $u$ is denoted by $b_u$ and it can be written as $b_u=b_v+b_w$. Each internal node is also associated with the IET between the last event in $v$ and the first event in $w$, denoted by $\hat\tau_u$. The index $u$ for the internal node follows the rank of its associated IET such that $\hat\tau_u\geq \hat\tau_{u'}$ for $u<u'$. For example, $u = 1$ for the root node as $\hat{\tau}_{1} = \tau_{\text{max}}$. Therefore, each internal node is characterized by a tuple of $(u,v,w,\hat{\tau}_{u})$, and the burst tree $\mathcal{T}$ is defined as the collection of these tuples, {$\{(u,v,w,\hat{\tau}_{u})\}$} for all nodes $u=1,\ldots,n-1$. We also remark that the burst tree $\mathcal{T}$ can be decomposed into the ordinal burst tree $\mathcal{G}=\{(u,v,w)\}$ and the set of IETs, i.e., $\{\hat{\tau}_{u}\}$ or $P(\tau)$. It implies that the ordinal burst tree can be associated with an arbitrary IET distribution other than the empirical IET distribution, while keeping the burst-tree structure intact. This is why several time series data from diverse backgrounds were found to show the similar burst-tree structure in terms of heavy-tailed burst size distributions and positive correlations between consecutive burst sizes for a wide range of timescales, despite system-specific IET distributions~\cite{Jo2020Bursttree}.

Since the burst tree contains exactly the same information in the event sequence, one can recover the original event sequence $\mathcal{E}$ from the derived burst tree $\mathcal{T}$ too. This is done by performing an inorder traversal of the burst tree. If the $i$th visited internal node is indexed by $u$, we denote it by $u(i)$. Then the event timings are recursively computed by $t_i=t_{i-1}+\hat\tau_{u(i)}$. Here $t_0$ can be taken from the time series data, or assumed to be $0$ for simplicity. Thanks to the derivability of the event sequence from the burst tree, one can devise a model generating an event sequence with a hierarchical structure of bursts; for this, we first generate an ordinal burst tree and then the event sequence by incorporating a set of IETs drawn from the IET distribution.

\section{Models and measures}\label{sec:model}

\subsection{Kernel-based modeling}\label{subsec:model}

Following Ref.~\cite{Jo2020Bursttree}, we take a kernel-based modeling approach to first generate the ordinal burst tree using several burst-merging kernels, and then event sequences by incorporating an interevent time (IET) distribution $P(\tau)$. The notion of the burst-merging kernel is based on the observation that as the timescale $\Delta t$ increases, bursts are merged to become bigger bursts (see Fig.~\ref{fig:burst_tree_fig}). Which bursts are merged together can be captured by the burst-merging kernel $K(b,b')$. That is, the probability of choosing two bursts of sizes $b$ and $b'$ for the merge is proportional to $K(b,b')$. We assume that this kernel is independent of $\Delta t$. In sum, the model has two inputs, namely, $K(b,b')$ and $P(\tau)$.

Let us begin with $n$ events or leaf nodes, each of which makes a burst of size one in the beginning. Their timings are to be calculated after the ordinal burst tree derived from the kernel is associated with the IET distribution. To describe the merging process in parallel with the coagulation process~\cite{Stockmayer1943Theory, Lushnikov1973Evolution, White1982Form, Hendriks1983Coagulation, Aldous1999Deterministic, Lee2001Survey, Leyvraz2003Scaling, Leyvraz2005Rigorous,Wattis2006Introduction}, we introduce an auxiliary time step $s$ that simply counts the number of merges in the merging process. We denote the burst size distribution at the time step $s$ by $Q_s(b)$. Then, at the initial time step $s=0$, one gets $Q_0(b)=\delta_{b,1}$, where $\delta$ is Kronecker delta. The merging process is as follows:
\begin{enumerate}
    \item At each time step $s$, randomly choose two bursts of sizes $b$ and $b'$ among $n-s$ bursts with a probability proportional to $K(b,b')$.
    \item Merge these two bursts to make another burst of size $b+b'$.
    \item The new burst becomes a parent node and two merged bursts are randomly assigned as left and right children of the parent node.
    \item Repeat (1--3) until $s$ reaches $n-1$, when a giant burst of size $n$ is formed, implying the root node in the ordinal burst tree.
\end{enumerate}
The burst size distribution at the final time step $s=n-1$ is obtained as $Q_{n-1}(b)=\delta_{b,n}$. Non-trivial behaviors of the burst size distributions are expected for $1\ll s\ll n-1$. Note that $Q_s(b)$ is directly derived from the ordinal burst tree at an auxiliary time step $s$, implying that it has nothing to do with real time.

We consider seven burst-merging kernels. The first three are constant, sum, and product kernels defined as follows:
\begin{align}
    &K_{\rm const}(b,b')=1, \label{eq:K_const}\\
    &K_{\rm sum}(b,b')=b+b', \label{eq:K_sum}\\
    &K_{\rm prod}(b,b')=bb'.\label{eq:K_prod}
\end{align}
Both $K_{\rm sum}$ and $K_{\rm prod}$ are increasing functions of $b$ and $b'$, implying that larger bursts are more likely to be chosen for the merge than smaller bursts. Therefore, these kernels indicate the preferential merging of bursts that is believed to be responsible for heavy-tailed burst size distributions. In contrast, since all three kernels do not require any similarity of bursts to be merged, they may not lead to the assortative pattern of bursts. Here the assortativity implies a tendency that bigger (smaller) bursts follow bigger (smaller) bursts~\cite{Jo2017Modeling}, accounting for positive correlations between consecutive burst sizes. To implement assortative merging of bursts, we consider a diagonal kernel as
\begin{align}
    K_{\rm diag}(b,b')=\delta_{b,b'}. \label{eq:K_diag}
\end{align}
With this kernel, only bursts of the same size can merge to become a bigger burst. Thus, one can expect the burst sizes to be in the form of $2^k$ for non-negative integers $k$. In addition, for the burst containing all events to exist, $n$ must be of the form of $2^k$.

The previous empirical results, e.g., using a Wikipedia editor's edit sequence, indicate both preferential and assortative merging of bursts, namely, the empirically estimated kernels $K(b,b')$ tend to increase with $b$ and $b'$ and to show higher values around the diagonal line of $b=b'$ than the off-diagonal regions~\cite{Jo2020Bursttree}. Inspired by such results, we here employ three more kernels dictating both preferential and assortative merging behaviors:
\begin{align}
    &K_{{\rm diag\text{-}pre}}(b,b')=b\, \delta_{b,b'}, \label{eq:K_diag-pre}\\
    &K_{{\rm sum\text{-}diag}}(b,b')= \frac{b+b'}{1+c_1(b-b')^2},\label{eq:K_sum-diag}\\
    &K_{\rm emp}(b,b') =\left[1 + c_2\ln (bb')\right] \left[1 + c_3e^{-c_4(\ln b - \ln b')^{2}}\right] \label{eq:K_emp},
\end{align}
where $c_1$, $c_2$, $c_3$, and $c_4$ are non-negative parameters. The diagonal-preferential kernel $K_{{\rm diag\text{-}pre}}$ is the diagonal kernel $K_{\rm diag}$ in Eq.~\eqref{eq:K_diag} multiplied by $b$, hence it is an increasing function of $b$ or $b'$ along the diagonal line $b=b'$. Next, setting $c_1=0$ in $K_{{\rm sum\text{-}diag}}$ reduces it to $K_{\rm sum}$; the denominator of $K_{{\rm sum\text{-}diag}}$ with positive $c_1$ implements the assortative merging of bursts. Finally, the first and second parentheses of $K_{\rm emp}$ indicate preferential and assortative merging, respectively. The functional complexity of $K_{\rm emp}$ is indeed inherited from the empirically observed kernel~\cite{Jo2020Bursttree}, hence we have suggested $K_{{\rm sum\text{-}diag}}$ as a simpler version of $K_{\rm emp}$. We also remark that all the kernels considered are symmetric with respect to their arguments, namely, $K(b,b')=K(b',b)$, implying the time-reversal symmetry of the generated ordinal burst tree.

Once the ordinal burst tree $\mathcal{G}$ is constructed, we then randomly draw $n-1$ IETs from $P(\tau)$ to get a set of $n-1$ IETs. We adopt a power-law IET distribution with the exponent $\alpha$, defined over the values of $\tau\in \{1,2,\ldots,\tau_{\rm c}\}$, as follows:
\begin{equation}
    P(\tau)=\frac{\tau^{-\alpha}}{\sum_{\tau=1}^{\tau_{\rm c}}\tau^{-\alpha}},
    \label{eq:Ptau}
\end{equation}
where $\tau_{\rm c}\gg 1$ is the upper bound of the IET. We assign each of the $n-1$ IETs, in a descending order from the largest to the smallest IETs, to the internal nodes, labeled from rank $1$ to rank $n-1$, respectively. Note that the root node with rank $1$ has the largest IET $\tau_{\text{max}}$. After setting $t_0=1$, we iteratively calculate the event timings using the recursive formula $t_{i} = t_{i-1} + \hat\tau_{u(i)}$, where $u(i)$ denotes the $i$th visited internal node when traversing the ordinal burst tree in inorder. 

\subsection{Measures for higher-order temporal correlations}\label{subsec:measure}

Using the generated event sequence we proceed to examine higher-order temporal correlations in terms of the burst size distribution $Q_{\Delta t}(b)$, memory coefficient for bursts $M_{\Delta t}$, and autocorrelation function (ACF) $A(t_{\rm d})$. For the first two measures, we take a horizontal cross-section of the burst tree at a particular timescale $\Delta t$, as shown by the dashed line in Fig.~\ref{fig:burst_tree_fig}. This cross-section yields a sequence of detected burst sizes; if $m$ bursts are detected, the sequence of burst sizes is given as $B_{\Delta t}=\{b_1,\ldots, b_m\}$. Such burst sizes can be summarized by the burst size distribution $Q_{\Delta t}(b)$, which often shows a power-law behavior with the exponent $\beta$ in several empirical analyses~\cite{Jo2020Bursttree} as
\begin{align}
    Q_{\Delta t}(b)\propto b^{-\beta}.
    \label{eq:Q_Deltat}
\end{align}
The burst size distribution however ignores any possible correlations between burst sizes. 

Correlations between burst sizes can be measured from $B_{\Delta t}$, e.g., in terms of the memory coefficient for bursts~\cite{Jo2020Bursttree}, which is defined as
\begin{align}
    M_{\Delta t} \equiv \frac{1}{m-1}\sum_{i = 1}^{m -1} \frac{(b_i - \mu_1)(b_{i+1} - \mu_2)}{\sigma_1 \sigma_2}.
    \label{eq:M_deltat}
\end{align}
Here $\mu_1$ ($\mu_2$) and $\sigma_1$ ($\sigma_2$) are the average and standard deviation of $B_{\Delta t}$ without the last (first) burst, respectively. $M_{\Delta t}$ has a value in the range of $[-1,1]$; $M_{\Delta t}$ may have a positive (negative) value if consecutive burst sizes are positively (negatively) correlated with each other. If the burst sizes are uncorrelated, $M_{\Delta t}$ is close to zero. 

\begin{figure*}[!t]
\centering
\includegraphics[width=0.8\textwidth]{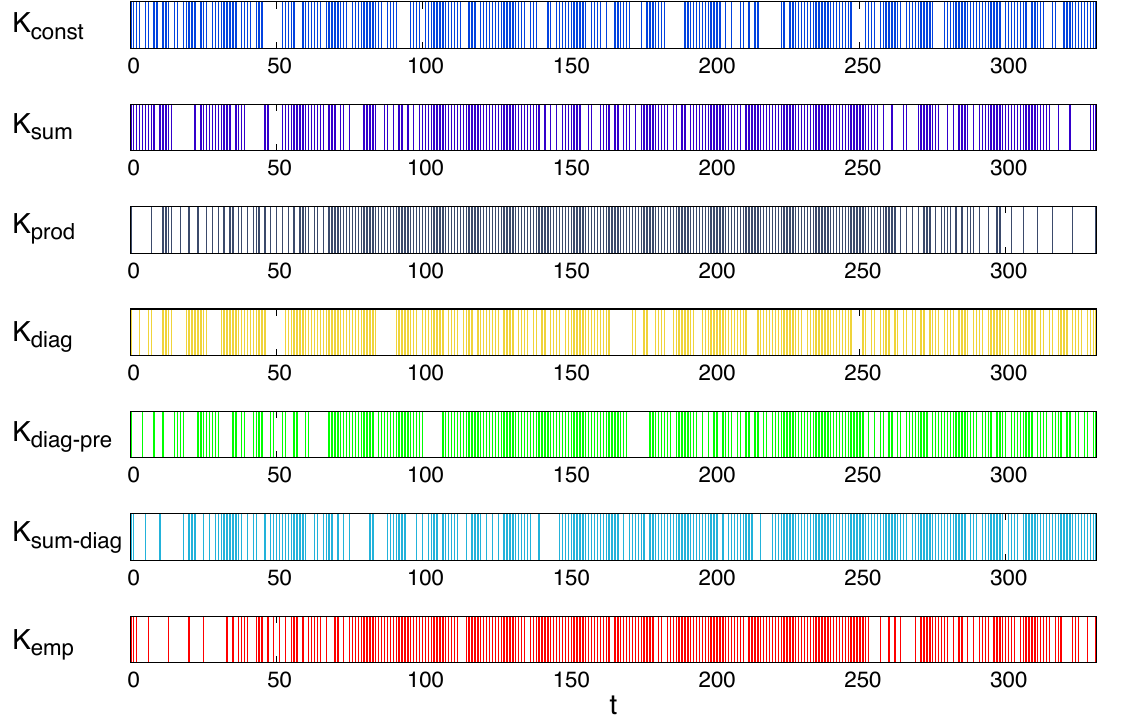}
\caption{Visualization of event sequences with $n=256$ events generated using the constant, sum, product, and diagonal kernels as well as kernels inspired by empirical results with $c_1=0.01$, $c_2=3$, $c_3=100$, and $c_4=0.25$ in Eqs.~\eqref{eq:K_const}--\eqref{eq:K_emp} (from top to bottom). Each vertical line depicts an event. We have used exactly the same set of 255 interevent times drawn from $P(\tau)\propto\tau^{-3.5}$ for all cases.}
\label{fig:time_series}
\end{figure*}

Finally, for the ACF, the event sequence is denoted by the time series $x(t)$ that has a value of $1$ if an event occurs at $t$, otherwise $0$. Assuming that the first event occurs at $t_0=1$, if the last event occurs at $t_{n-1}=T$, the ACF with a time lag $t_{\rm d}$ is calculated as
\begin{align}
    A(t_{\rm d})\equiv 
    \frac{\frac{1}{T- t_{\rm d}}\sum_{t = 1}^{T-t_{\rm d}}x(t)x(t+t_{\rm d}) - \mu^x_1\mu^x_2}{\sigma^x_1\sigma^x_2},
    \label{eq:acf}
\end{align}
where $\mu^x_1$ and $\sigma^x_1$ denote the average and standard deviation of $\{x(1),\ldots,x(T-t_{\rm d})\} $, respectively, and $\mu^x_2$ and $\sigma^x_2$ denote the average and standard deviation of $\{x(t_{\rm d}+1),\ldots,x(T)\}$, respectively. When there are long-range memory effects in the time series, the ACF tends to decay algebraically with the decay exponent $\gamma$ as
\begin{align}
    A(t_{\rm d})\propto t_{\rm d}^{-\gamma}.
    \label{eq:gamma}
\end{align}
It is expected that $\gamma$ is dependent on the kernel $K(b,b')$ and the IET exponent $\alpha$~\cite{Lowen1993Fractal, Jo2024Temporal}.

\subsection{Analytical results of burst size distributions}

Thanks to the analogy between the merging process and the coagulation process, exact solutions of $Q_s(b)$ are available for the constant, sum, and product kernels~\cite{Wattis2006Introduction}. In the case of the constant kernel, the cluster size distribution in the coagulation process translates into the burst size distribution as follows:
\begin{align}
    Q_s(b)\simeq \frac{1}{b_s} e^{-b/b_s},
    \label{eq:wattis_const}
\end{align}
where $b_s\equiv \frac{t(s)}{2}$. Here the time variable $t$ in the coagulation process is an increasing function of $s$ in the merging process but its functional form is not always trivial to obtain. Since $b_s$ in Eq.~\eqref{eq:wattis_const} is the same as the average burst size, which is $\frac{n}{n-s}$, one gets $t(s)=\frac{2n}{n-s}$. The exponential burst size distribution implies that there is no correlation between IETs~\cite{Karsai2012Universal}. In the case of the sum kernel, we can write
\begin{align}
    Q_s(b) \propto b^{-3/2}e^{-b/b'_s},
    \label{eq:wattis_add}
\end{align}
where $b'_s \equiv 2e^{t(s)}$. Again, $t$ in the coagulation process is a function of $s$ in the merging process. The burst size distribution shows a power-law regime with exponent $\beta=3/2$. Next, in the case of the product kernel, we get
\begin{align}
    Q_s(b) \propto \begin{cases}
    b^{-5/2}e^{-b/b''_s} &\textrm{for}\ t(s) < 1, \\
    b^{-5/2} &\textrm{for}\ t(s) \geq 1,
    \label{eq:wattis_product}
  \end{cases}
\end{align}
where $b''_s\equiv \frac{1}{t(s)-1-\log t(s)}$. In the context of the coagulation process, a gelation transition occurs at $t(s)=1$, when a particle of infinite size or a gel appears~\cite{Wattis2006Introduction}. This implies in the merging process that a very big burst emerges at a finite time step $s$, satisfying $t(s)=1$. The burst size distribution shows a power-law regime with exponent $\beta=5/2$ for $t(s)<1$, while it has a power-law tail with the same exponent for $t(s)\geq 1$. 

Finally, the diagonal-preferential kernel $K_{{\rm diag\text{-}pre}}$ can be considered as a special case of the form of $K(x,y)=x^{1+\lambda}\delta(x-y)$ that was studied in Ref.~\cite{Leyvraz2005Rigorous}, where $\lambda\geq 0$, and $x$ and $y$ are continuous variables. The analytical solution of the cluster size distribution for the case with $\lambda=0$ in Ref.~\cite{Leyvraz2005Rigorous} can be translated into the burst size distribution as
\begin{align}
    Q_s(b) \propto \frac{b^{-1}}{t(s)}\left[1-c\left(\frac{b}{t(s)}\right)^{\Delta}\right],
    \label{eq:leyvraz_diag}
\end{align}
where $c>0$ is a constant determined by the normalization, and $\Delta\approx 2.69$. These analytical results will be compared to the simulation results in the next Section, but indirectly via $Q_{\Delta t}(b)$. It is because once the number of bursts for a given $\Delta t$ is determined as $m_{\Delta t}$, the corresponding auxiliary time step $s$ is simply given as $n-m_{\Delta t}$. Then, since $Q_{s=n-m_{\Delta t}}(b)=Q_{\Delta t}(b)$, it is enough to show the simulation results of $Q_{\Delta t}(b)$ to see if the analytic solutions predict the simulations well.

\begin{figure*}[!t]
\centering
\includegraphics[width=0.8\textwidth]{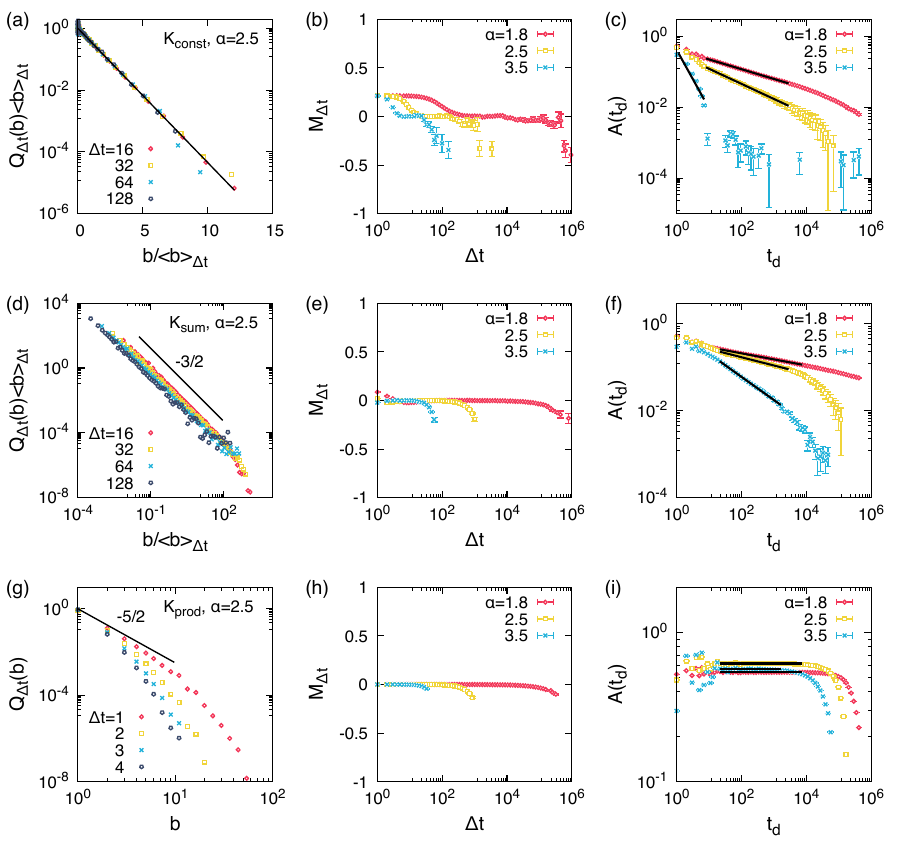}
\caption{Simulation results of burst size distributions $Q_{\Delta t}(b)$, memory coefficients for bursts $M_{\Delta t}$ in Eq.~\eqref{eq:M_deltat}, and autocorrelation functions $A(t_{\rm d})$ in Eq.~\eqref{eq:acf} (from left to right) obtained from $50$ event sequences of $n = 5\cdot 10^5$ events generated using $K_{\rm const}$, $K_{\rm sum}$, and $K_{\rm prod}$ [Eqs.~\eqref{eq:K_const}--\eqref{eq:K_prod}] (from top to bottom). In panels (a,~d,~g), we show $Q_{\Delta t}(b)$ for several values of $\Delta t$ in the case with the IET exponent $\alpha=2.5$, and $\langle b\rangle_{\Delta t}$ is the average burst size of $Q_{\Delta t}(b)$. Black lines without or with exponent values in panels (a,~d,~g) are for guiding eyes. Black lines in panels (c,~f,~i) show fitting ranges and results of the decay exponent $\gamma$ in Eq.~\eqref{eq:gamma}, summarized in Table~\ref{table:gamma}. Error bars in $M_{\Delta t}$ and $A(t_{\rm d})$ denote standard errors.}
\label{fig:summary1}
\end{figure*}

\section{Numerical results}\label{sec:result}

To demonstrate the effects of kernels on higher-order temporal correlations, we visualize in Fig.~\ref{fig:time_series} the entire event sequences of $n=256$ events generated using seven kernels in Eqs.~\eqref{eq:K_const}--\eqref{eq:K_emp}. For emphasizing the role of burst tree structure, we use exactly the same set of 255 interevent times (IETs) drawn from $P(\tau)$ in Eq.~\eqref{eq:Ptau} with $\alpha=3.5$ and $\tau_{\rm c}=10^7$ for all cases. As expected, the event sequence using $K_{\rm const}$ shows no higher-order temporal correlations, compared to other event sequences showing such correlations. In the event sequence using $K_{\rm sum}$, which is expected to lead to the heavy-tailed burst size distribution, one can observe bigger bursts than those in the case with $K_{\rm const}$, such as bursts in the interval of $190\lesssim t\lesssim 260$. The event sequence using $K_{\rm prod}$ obviously shows a very big burst corresponding to the gel formed in the coagulation process. The other four event sequences are generated kernels having the feature of assortative merging; all of them show not only bigger bursts than those in the case with $K_{\rm const}$, but also bursts of similar sizes following each other, which are not expected in the cases of constant, sum, and product kernels. Such an assortative pattern is evidenced most clearly in the case with $K_{\rm diag}$. It is also seen for the case with $K_{{\rm sum\text{-}diag}}$, e.g., in the intervals of $10\lesssim t\lesssim 50$ and $80\lesssim t\lesssim 130$, where relatively small bursts are followed by similarly small bursts. Finally, we note that although $K_{\rm diag}$ does not implement the preferential merging by definition, the generated event sequence shows some bigger bursts than those by $K_{\rm const}$, possibly implying that the heavy-tailed burst size distributions may be observed due to the assortative merging only. This issue will be discussed later.

For the systematic comparison, we perform large-scale numerical simulations by generating $50$ different event sequences for each combination of seven kernels in Eqs.~\eqref{eq:K_const}--\eqref{eq:K_emp} and the IET exponents $\alpha=1.8$, $2.5$, and $3.5$ of the IET distribution with $\tau_{\rm c}=10^7$ in Eq.~\eqref{eq:Ptau}. The number of events in each event sequence is $n=5\cdot 10^5$ for all kernels except for $K_{\rm diag}$ and $K_{{\rm diag\text{-}pre}}$, for which $n=2^{19}=524,288$ has been used. Using the generated event sequences we measure burst size distributions $Q_{\Delta t}(b)$ for several values of $\Delta t$, memory coefficients for bursts $M_{\Delta t}$ for a wide range of $\Delta t$ in Eq.~\eqref{eq:M_deltat}, and the autocorrelation function (ACF) $A(t_{\rm d})$ in Eq.~\eqref{eq:acf}.

Let us first look at the cases with constant, sum, and product kernels as shown in Fig.~\ref{fig:summary1}. The burst size distributions at several timescales, $Q_{\Delta t}(b)$, are exponential for $K_{\rm const}$ as expected. For $K_{\rm sum}$ and $K_{\rm prod}$, $Q_{\Delta t}(b)$s at several values of ${\Delta t}$ show power-law scaling regimes with exponents $3/2$ and $5/2$, respectively. Note that for the case with $K_{\rm prod}$, since almost all events make very large bursts as partly shown in Fig.~\ref{fig:time_series}, the rest of events may not be enough to fully develop the power-law tails in the burst size distributions. Such large bursts correspond to the gels in the coagulation process and they appear in the burst size distributions as small peaks at the very large $b$ of the order of $n$, which are not shown in Fig.~\ref{fig:summary1}(g).

As mentioned, all of constant, sum, and product kernels may not lead to positive correlations between consecutive burst sizes. Indeed, we observe $M_{\Delta t}\approx 0$ for these kernels. However, in the case of $K_{\rm const}$, we find $M_{\Delta t}\approx 0.2$ for small values of $\Delta t$ [Fig.~\ref{fig:summary1}(b)]. To explain this observation, we remind that for small $\Delta t$, there are a number of bursts of sizes $1$ or $2$. If these bursts were located in the sequence of burst sizes in a random manner, $M_{\Delta t}$ would be negligible. Instead, bursts of size $2$ seem to be located next to each other more often than expected, possibly due to the rule of randomly assigning left and right children nodes to the merged bursts. For example, if four bursts of sizes $1$, $1$, $2$, and $2$ are merged to two bursts of size $3$, which in turn are merged to a burst of size $6$, the sequence of burst sizes of $\{1,2,2,1\}$ may happen with a probability of $1/4$, eventually resulting in the positive $M_{\Delta t}$. This general argument however does not apply to the cases with $K_{\rm sum}$ and $K_{\rm prod}$ because the preferential merging forces consecutive burst sizes to differ from each other very quickly, leading to the negligible values of $M_{\Delta t}$. The negative values of $M_{\Delta t}$ for a very large $\Delta t$ are due to the finite-size effects as a few very big bursts and other relatively small bursts remain for such a large $\Delta t$. We finally remark that the IET exponent $\alpha$ affects only the timescales of the burst tree $\mathcal{T}$ not the ordinal burst tree $\mathcal{G}$; the smaller value of $\alpha$ implies a wider range of IETs and timescales. Therefore, the curve of $M_{\Delta t}$ for smaller $\alpha$ can be seen as a stretched version of that for larger $\alpha$.

\begin{table}[!t]
\caption{Estimated values and fitting ranges of the decaying exponent $\gamma$ of ACFs presented in Figs.~\ref{fig:summary1} and~\ref{fig:summary2} for several values of the IET exponent $\alpha$ per kernel used. ``P'' and ``A'' mean the preferential and assortative merging rules, respectively. Values of the burst size exponent $\beta$ are exact for the constant, sum, and product kernels~\cite{Wattis2006Introduction}, while for the other kernels, values of $\beta$ are estimated from the numerical results of $Q_{\Delta t=64}(b)$. Numbers in parentheses are errors.}
\centering 
\begin{tabular}{lccccccc} 
\hline \hline 
Kernel & P & A & $\beta$ & $\alpha$ & & $\gamma$ & Fitting range of $t_{\rm d}$ \\ 
\hline
$K_{\rm const}$ & $\times$ & $\times$ & $\infty$ & 1.8 & & 0.28(1) & [$e^2$, $e^8$] \\
& & & & 2.5 & & 0.43(1) & [$e^2$, $e^8$] \\
& & & & 3.5 & & 1.6(2) & [$e^0$, $e^2$] \\
\hline
$K_{\rm sum}$ & \checkmark & $\times$ & $3/2$ & 1.8 & & 0.14(1) & [$e^3$, $e^9$] \\
& & & & 2.5 & & 0.19(1) & [$e^3$, $e^8$] \\
& & & & 3.5 & & 0.53(1) & [$e^3$, $e^{7.5}$] \\
\hline 
$K_{\rm prod}$ & \checkmark & $\times$ & $5/2$ & 1.8 & & 0.00(1) & [$e^3$, $e^9$] \\
& & & & 2.5 & & 0.00(1) & [$e^3$, $e^9$] \\
& & & & 3.5 & & 0.00(1) & [$e^3$, $e^{7.5}$] \\
\hline 
$K_{\rm diag}$ & $\times$ & \checkmark & 1.00(1)\footnote{Fitted for $b\in [e^{2.5}, e^{8.5}]$.} & 1.8 & & 0.23(1) & [$e^3$, $e^{12}$] \\
& & & & 2.5 & & 0.36(1) & [$e^2$, $e^9$] \\
& & & &  3.5 & & 0.77(7) & [$e^2$, $e^8$] \\
\hline 
$K_{{\rm diag\text{-}pre}}$ & \checkmark & \checkmark & 0.85(2)\footnote{Fitted for $b\in [e^{2.5}, e^{7}]$.} & 1.8 & & 0.089(1) & [$e^3$, $e^9$] \\
& & & & 2.5 & & 0.14(1) & [$e^3$, $e^8$] \\
& & & & 3.5 & & 0.33(1) & [$e^3$, $e^6$] \\
\hline 
$K_{{\rm sum\text{-}diag}}$ & \checkmark & \checkmark & 1.02(2)\footnote{Fitted for $b\in [e^{2}, e^{7.5}]$.} & 1.8 & & 0.19(1) & [$e^2$, $e^7$] \\
& & & & 2.5 & & 0.20(1) & [$e^3$, $e^8$] \\
& & & & 3.5 & & 0.38(1) & [$e^2$, $e^7$] \\
\hline 
$K_{\rm emp}$ & \checkmark & \checkmark & 1.28(2)\footnote{Fitted for $b\in [e^{2}, e^{7.5}]$.} & 1.8 & & 0.14(1) & [$e^2$, $e^{10}$] \\
& & & & 2.5 & & 0.16(1) & [$e^2$, $e^9$] \\
& & & & 3.5 & & 0.29(1) & [$e^2$, $e^8$] \\
\hline \hline
\end{tabular}
\label{table:gamma} 
\end{table}

\begin{figure*}[!t]
\centering
\includegraphics[width=0.8\textwidth]{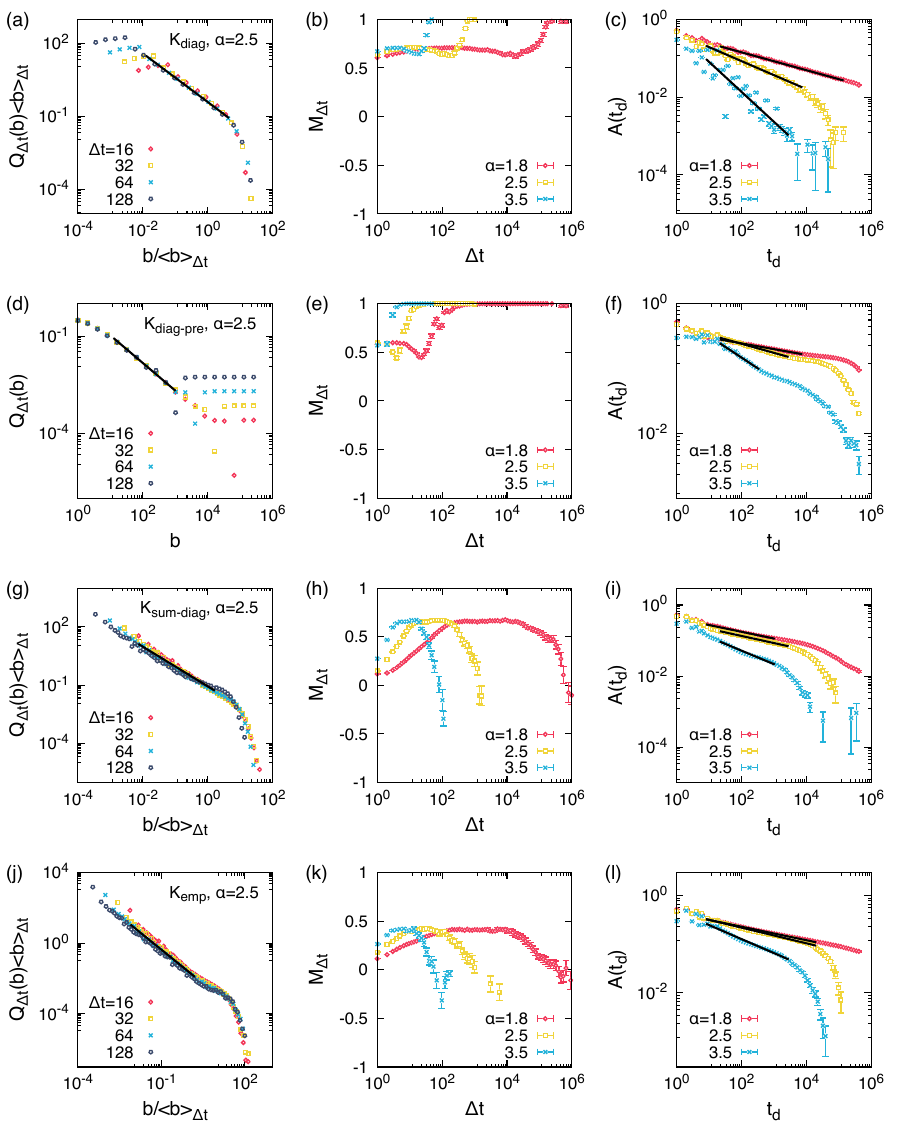}
\caption{The same as Fig.~\ref{fig:summary1} but using $K_{\rm diag}$, $K_{{\rm diag\text{-}pre}}$, $K_{{\rm sum\text{-}diag}}$ with $c_1=0.01$, and $K_{\rm emp}$ with $c_2=3$, $c_3=100$, and $c_4=0.25$ [Eqs.~\eqref{eq:K_diag}--\eqref{eq:K_emp}] (from top to bottom). For $K_{\rm diag}$ and $K_{{\rm diag\text{-}pre}}$, we use $n=2^{19}=524,288$. Black lines in panels (a,~d,~g,~j) show fitting ranges and results of the burst-size exponent $\beta$ in Eq.~\eqref{eq:Q_Deltat}, also summarized in Table~\ref{table:gamma}. 
}
\label{fig:summary2}
\end{figure*}

In Fig.~\ref{fig:summary1}(c,~f,~i), we find that the ACFs decay algebraically with the time lag $t_{\rm d}$. We estimate the value of the decay exponent $\gamma$ by the ordinary least squares linear regression using the equation of $\ln A(t_{\rm d})=a_0-\gamma \ln t_{\rm d}$ with a constant $a_0$ for the scaling regime of $t_{\rm d}$. The results are summarized in Table~\ref{table:gamma} with the fitting ragne of $t_{\rm d}$. The event sequences generated using $K_{\rm const}$ can be considered as results of the renewal process, in which there is no correlations between IETs. In that case, $\gamma$ has been known to be a function of $\alpha$ only; precisely, $\gamma=2-\alpha$ for $1<\alpha<2$ and $\gamma=\alpha-2$ for $\alpha>2$~\cite{Lowen1993Fractal}. However, numerical studies have shown strong finite-size effects on the mentioned scaling relation~\cite{Jo2017Modeling}. Considering such finite-size effects, the estimated values of $\gamma$ in Table~\ref{table:gamma} seem to be overall consistent with the analytical results of $\gamma$. In the case of $K_{\rm sum}$, $\beta=3/2$ implies long-range temporal correlations between consecutive IETs, leading to different values of $\gamma$ than those for $K_{\rm const}$. Although there are no analytical solutions of the ACF for $K_{\rm sum}$, one can refer to the recent analytical result of $\gamma$ as a function of $\alpha$ and $\beta$ in Ref.~\cite{Jo2024Temporal}; for the range of $1<\beta<2$, $\gamma$ is an increasing function of $\alpha$. Thus, it is qualitatively consistent with the numerical values of $\gamma$ for $K_{\rm sum}$ in Table~\ref{table:gamma}. We remark that since the model studied in Ref.~\cite{Jo2024Temporal} considers only one timescale $\Delta t$, its analytical results might systematically deviate from numerical results of our model considering the entire range of $\Delta t$.

The ACFs for $K_{\rm prod}$ are found to be constant for a wide range of $t_{\rm d}$ as shown in Fig.~\ref{fig:summary1}(i), indicating $\gamma\approx 0$. One can get a hint from the visualized event sequence in Fig.~\ref{fig:time_series}, in which a very big burst containing almost all events is formed at a small value of $\Delta t$. The similar flat behaviors of the ACF have been reported in a model study based on the self-exciting point process~\cite{Jo2015Correlated}, where the events form very big bursts that are separated by very large IETs (see Fig.~7 in Ref.~\cite{Jo2015Correlated}).

Simulation results using the other four kernels are presented in Fig.~\ref{fig:summary2}. We first discuss the case with the diagonal kernel $K_{\rm diag}$ in Eq.~\eqref{eq:K_diag}; although it does not have a feature for the preferential merging, the power-law regime with $\beta\approx 1$ is clearly observed in the burst size distributions. This indicates that the heavy-tailed burst size distributions might be not only due to the preferential merging but also due to the assortative merging. To understand this observation, one can map the merging process to a system of particles on one-dimensional lattice, where each site $i$ corresponds to the burst size $b=2^i$ ($i=0,1,\ldots$), all particles are initially at the site $0$, and two particles at the same site $i$ hop to the next site $i+1$ after merging. Since the probability of choosing two particles is constant irrespective of their position, the number of particles at the site $i$ can be written as $n_i\propto c^i$ with a constant $c\in (0,1)$, which is combined with $b=2^i$ to give $Q(b)\propto b^{-\beta}$ with $\beta=-\log_2 c>0$. We also find that $M_{\Delta t}>0$ for the entire range of $\Delta t$, and that $\gamma$ increases with $\alpha$, again qualitatively consistent with the analytical result in Ref.~\cite{Jo2024Temporal}.

Finally, we find that the other kernels, i.e., $K_{{\rm diag\text{-}pre}}$, $K_{{\rm sum\text{-}diag}}$, and $K_{\rm emp}$, which implement both preferential and assortative merging, lead to heavy-tailed burst size distributions and positive correlations between consecutive burst sizes at the same time, as expected. From now on we focus on the scaling behaviors of $Q_{\Delta t}(b)\propto b^{-\beta}$ and $A(t_{\rm d})\propto t_{\rm d}^{-\gamma}$ because $M_{\Delta t}>0$ is naturally expected in all cases.

The estimated value of $\beta$ for $K_{{\rm diag\text{-}pre}}$ is $\approx 0.85$, indicating a heavier tail than that for $K_{\rm diag}$. This can be understood by considering the fact that $K_{{\rm diag\text{-}pre}}$ has a preferential factor $b$ multiplied to $K_{\rm diag}$. However, such a numerical result is not consistent with the analytical solution in Eq.~\eqref{eq:leyvraz_diag}, possibly due to the fact that the analytical solution was obtained for continuous variables of masses in the coagulation process in contrast to the discrete nature of the burst-merging process. This requires further investigation in the future. Next, by comparing $\beta\approx 1$ for $K_{{\rm sum\text{-}diag}}$ to $\beta=3/2$ for $K_{\rm sum}$, we observe that the tails of burst size distributions get heavier due to the assortative merging with $c_1>0$ in Eq.~\eqref{eq:K_sum-diag}. We also find $\beta\approx 1.28$ for $K_{\rm emp}$. It turns out that estimated values of $\beta$ for these three kernels are smaller than $2$, allowing us to expect that $\gamma$ increases with $\alpha$~\cite{Jo2024Temporal}, which is indeed the case as shown in Table~\ref{table:gamma}. The scaling behaviors for the case with $K_{\rm emp}$ appear most similar to those with $K_{\rm sum}$, except for the value of $\gamma$ when $\alpha=3.5$ is used.

\section{Conclusion}\label{sec:conclusion}

We have studied how different burst-merging kernels shape the higher-order temporal correlations in the burst tree structure as well as the time series. For this, we adopt constant, sum, product, and diagonal kernels as well as those inspired by the previous empirical results~\cite{Jo2020Bursttree}. To characterize the higher-order temporal correlations, we measure the burst size distributions, memory coefficients for bursts, and the autocorrelation function. By large-scale numerical simulations, we generically find that kernels with preferential merging lead to the heavy-tailed burst size distributions, while kernels with assortative merging lead to positive correlations between burst sizes. We also find that in some kernels the assortative merging enhances the heavy tail in burst size distributions. The decaying exponent of the autocorrelation function depends not only on the kernel but also on the power-law exponent of the interevent time distribution. Therefore, our findings would shed lights on understanding various empirical results reported in Ref.~\cite{Jo2020Bursttree}. 

Although we have focused on the preferential and assortative mixing structure of bursts, it is also possible that other kinds of features in burst-merging kernels may be observed in other empirical time series. Then our understanding of the simple kernels may serve as a reference to understand such novel features and their impact on the higher-order temporal correlations in such real-world phenomena.

Finally, despite the usefulness of our framework using burst-merging kernels in generating or analyzing higher-order temporal correlations, underlying mechanisms behind such kernels are not yet fully understood. The kernels are used to determine which bursts are merged to become bigger bursts, rather than explain the reasons for such a determination at the most fundamental level. Thus, it would be interesting if one could devise more mechanistic processes, e.g., based on queuing models~\cite{Barabasi2005Origin, Vazquez2006Modeling, Jo2012Timevarying} or self-exciting point processes~\cite{Adamopoulos1976Cluster, Utsu1995Centenary, Ogata2006Space, Jo2015Correlated}, and translate them into the burst-merging kernels. By doing so, one may relate elements in the mechanistic processes to the features of corresponding kernels.

\begin{acknowledgments}
H.-H.J. thanks Mikko Kivel\"a and Takayuki Hiraoka for fruitful discussions. T.B. and H.-H.J. acknowledge financial support by the National Research Foundation of Korea (NRF) grant funded by the Korea government (MSIT) (No. 2022R1A2C1007358). This work was supported by the Research Fund of The Catholic University of Korea in 2024.
\end{acknowledgments}




\bibliography{paper}

\end{document}